\documentclass[twocolumn,aps,prb]{revtex4-1}
\usepackage{graphicx}
\usepackage{amsmath, bm}

\newcommand{\fn}{\sigma}
\newcommand{\sn}{{\sigma}}

\newcommand{\cH}{{\cal H}}
\newcommand{\cV}{{\cal V}}

\newcommand{\cHel}{\cH_\text{el}}

\newcommand{\rt}{t}

\newcommand{\eU}{U_{\mbox{\scriptsize eff}}}
\newcommand{\sthermo}{{\mbox{\scriptsize thermo}}}

\newcommand{\Eg}{E_{\mbox{\scriptsize gap}}}

\newcommand{\prtl}{\partial}

\begin{document}

\newlength{\figurewidth}
\setlength{\figurewidth}{\columnwidth}

\title{Electronic structure and the glass transition in pnictide and
  chalcogenide semiconductor alloys. Part II: The intrinsic electronic
  midgap states }
\author{Andriy Zhugayevych$^1$}
\author{Vassiliy Lubchenko$^{1,2}$}\email{vas@uh.edu}
\affiliation{$^1$Department of Chemistry, University of Houston, TX
  77204-5003 \\ $^2$Department of Physics, University of Houston, TX
  77204-5005} \date{\today}
\begin{abstract}

  We propose a structural model that treats in a unified fashion both
  the atomic motions and electronic excitations in quenched melts of
  pnictide and chalcogenide semiconductors. In Part I (submitted to
  {\it J. Chem. Phys.}), we argued these quenched melts represent
  aperiodic $pp\sigma$-networks that are highly stable and, at the
  same time, structurally degenerate. These networks are characterized
  by a \emph{continuous} range of coordination.  Here we present a
  systematic way to classify these types of coordination in terms of
  \emph{discrete} coordination defects in a parent structure defined
  on a simple cubic lattice. We identify the lowest energy
  coordination defects with the intrinsic midgap electronic states in
  semiconductor glasses, which were argued earlier to cause many of
  the unique optoelectronic anomalies in these materials. In addition,
  these coordination defects are mobile and correspond to the
  transition state configurations during the activated transport above
  the glass transition. The presence of the coordination defects may
  account for the puzzling discrepancy between the kinetic and
  thermodynamic fragility in chalcogenides. Finally, the proposed
  model recovers as limiting cases several popular types of bonding
  patterns proposed earlier, including: valence-alternation pairs,
  hypervalent configurations, and homopolar bonds in heteropolar
  compounds.

\end{abstract}


\maketitle

\section{Introduction}

In the preceding article,\cite{ZLMicro1} we presented a chemical
bonding theory for an important class of pnictogen- and
chalcogen-containing quenched melts and glasses. These materials
exhibit many unique electronic and optical
anomalies\cite{ShimakawaElliott} not found in crystals, and also are
of great interest in applications such as information storage and
processing.\cite{ISI:000250615400019, ISI:000261127100019} We argued
these materials can be thought of as aperiodic $pp\sigma$-networks
made of deformed-linear chains that intersect at atomic sites at
nearly right angles. Extended portions of the chains exhibit a perfect
alternation of covalent and weaker, secondary bonds, even though the
lattice as a whole is aperiodic. This bond alternation results from a
symmetry breaking of a parent, simple cubic structure with octahedral
coordination; this structure is uniformly covalently bonded. It is the
intimate relation of the amorphous lattice to its covalently bonded
parent structure that allowed us to rationalize, for the first time to
our knowledge, two seemingly contradicting features of a bulk glass,
i.e., its relative stability and structural
\emph{degeneracy}.\cite{ZLMicro1}

Yet the argued presence of the degeneracy of the $pp\sigma$-network
does not, by itself, guarantee that the network can be realized as an
equilibrated supercooled liquid or a quenched glass: For instance,
elemental arsenic can be made into an amorphous film but does not
vitrify readily.  To be a liquid, the network should contain a large
\emph{equilibrium} concentration of structural motifs corresponding to
the transition states for activated transport in quenched melts.

The goal of the present article is to identify the bonding patterns of
the transition-state structural motifs at the molecular level and
describe the rather peculiar midgap electronic states that are
intrinsically associated with such motifs. It will turn out that these
electronic states are responsible for many of the aforementioned
electronic and optical anomalies of amorphous chalcogenide alloys.

Identification of structural motifs in vitreous materials based on
local coordination is difficult because the usual concept of
coordination, which is not fully unambiguous even in periodic
lattices, becomes even less compelling in aperiodic
systems. Alternatively, one can try to classify such local motifs in
terms of deviation from a putative reference structure, while
assigning a corresponding energy cost. Such deviations could be called
``defected configurations.''  However, in the absence of long-range
order, defining a reference structure in vitreous systems is, again,
ambiguous. To make an informal analogy, is there a way to identify a
typo in a table of random numbers? The answer, of course, depends on
the presence and specific type of correlation between the random
numbers.  The random first order transition (RFOT)
theory\cite{LW_ARPC} dictates that glasses do form subject to strict
statistical rules prescribed by the precise degree of structural
degeneracy of the lattice, implying that correlation functions of
sufficiently high order should reveal defects, if any.  In fact,
already four-point correlation functions in space capture the length
scale of the dynamic heterogeneity in quenched melts.\cite{Berthier,
  Dalle-Ferrier, Capaccioli} Yet the only type of order in glasses
that appears to be unambiguously accessible to \emph{linear}
spectroscopy is the very shortest-range order: The very first
coordination layer is usually straightforward to identify by
diffraction experiments, while the strong covalent bonding between
nearest neighbors is identifiable via the independent knowledge of the
covalent radii of the pertinent elements. Already the next-nearest
neighbor bonds appear to exhibit a \emph{continuous} range of strength
and mutual angular orientation.

We have argued\cite{ZLMicro1} aperiodic $pp\sigma$-networks naturally
account for this flexibility in bonding in semiconductor glasses,
while retaining overall stability. In doing so, we proposed a
structural model, by which aperiodic $pp\sigma$-networks can be
thought of as distorted versions of much simpler \emph{parent}
structures defined on the simple cubic lattice. There is no ambiguity
whatsoever with defining coordination on a simple cubic lattice, thus
allowing one to classify unambiguously the parent structures. 

We will observe that the most important and essentially the sole type
of defect in $pp\sigma$-bonded glasses is singly over- or
under-coordinated atoms. These defects turn out to host peculiar
midgap electronic states with the reversed charge-spin relation, i.e.,
chemically they resemble free radicals. We will argue that these
defects in fact correspond to the electronic states residing on the
high-strain regions intrinsic to the activated transport in
semiconductor glasses proposed earlier by us.\cite{ZL_JCP} On the one
hand, these electronic states lie very deep in the forbidden gap. On
the other hand, they are surprisingly extended, calling into question
the adequacy of ultralocal defect models. This large spatial extent
reveals itself by redistribution of the malcoordination over a large
number of bonds, in a solitonic fashion, and delocalization of the
wave function of the associated electronic state. The extended
coordination defects are mobile, consistent with the conclusions of
our earlier semi-phenomenological analysis\cite{ZL_JCP} that the
peculiar electronic states are hosted by high-strain regions that
emerge during activated transport in quenched semiconductor melts.

The article is organized as follows. Section \ref{RFOT} reviews the
conclusions of the RFOT theory on the concentration and spatial
characteristics of the transition state configurations for activated
transport in supercooled liquids and frozen glasses, and a general
mechanism for the emergence of associated electronic
states.\cite{ZL_JCP} In Section \ref{parent}, a systematic
classification of coordination defects in parent structures is carried
out. In Section \ref{soliton}, we demonstrate that malcoordination
defects in parent structures become delocalized in the actual, relaxed
structure and make a connection with our earlier,
semi-phenomenological analysis of the solitonic states.\cite{ZL_JCP}
When making this connection, we perform several independent
consistency checks that the degeneracy of aperiodic
$pp\sigma$-networks is indeed compatible with the degeneracy of actual
semiconductor alloys. Lastly we argue that the electronic states make
a temperature-independent contribution to the activation barrier for
liquid reconfigurations, which helps explain the apparent disagreement
between the thermodynamic and kinetic fragilities in chalcogenides.

We have alluded to two vulnerabilities of ad hoc defect theories,
i.e., the difficulty in defining a putative reference structure and
the presumption of defects' being ultralocal at the onset. Conversely,
such ad hoc theories do not explain self-consistently how the defects
combine to form an actual, quite stable lattice. The present approach
resolves this potential ambiguity. First, since the defects are
defined on a specific lattice in the first place, the question of
their coexistence in a 3D structure is automatically answered.
Second, as discussed in Section \ref{sec.model}, we will see how
several popular defect theories proposed much earlier on
phenomenological grounds are naturally recovered as the ultralocal
limit of the present picture.

\section{Brief Review of the RFOT theory and Previous Work}
\label{RFOT}

Below we outline the minimum set of notions from the Random First
Order Transition (RFOT) theory of the glass transition and other
previous work, as necessary for the subsequent developments. Detailed
reviews of the RFOT theory can be found elsewhere.\cite{LW_ARPC,
  LW_RMP}

Mass transport in liquids slows down with lowering temperature or
increasing density because of increasingly more frequent molecular
collisions.  At viscosities of $~10$ Poise and above, however, the
transport is no longer dominated by collisions but, instead, is in a
distinct dynamical regime: The equilibrium liquid density
profile\cite{Evans1979} is no longer uniform, but instead consists of
sharp disparate peaks,\cite{dens_F1} whereby each atom \emph{vibrates}
around a fixed location in space for an extended period of time. In
other words, the liquid is essentially an assembly of long-living
structures that persist for times exceeding the typical relaxation
times of the vibrations by at least three orders of
magnitude.\cite{LW_soft} Under these circumstances, mass transport
becomes activated: Atoms move cooperatively via barrier-crossing
events whereby the current long-lived, low free energy aperiodic
configuration transitions locally to another long-lived, low free
energy configuration. The multiplicity of alternative aperiodic
configurations is quantified using the so called configurational
entropy: A region of size $N$ particles has $e^{s_c N/k_B}$ distinct
structural states, where $s_c$ is the configurational entropy per per
rigid group of atoms, often called the ``bead.''\cite{LW_soft} The
RFOT theory predicts that the configurational entropy at the glass
transition is $\simeq 0.8 k_B$ per bead.\cite{XW}

Individual atomic displacements during the transitions are small,
i.e., about the vibrational displacement at the mechanical stability
edge, which is often called the Lindemann length
$d_L$.\cite{Lindemann} Although the activated transport regime is
usually associated with supercooled liquids,\cite{app_phys_rev, tenQ}
many covalently bonded substances, such as SiO$_2$, exhibit activated
transport already above the melting point.\cite{LW_soft}

\begin{figure}[t] 
      \begin{center}
        \includegraphics[width= .8 \figurewidth]{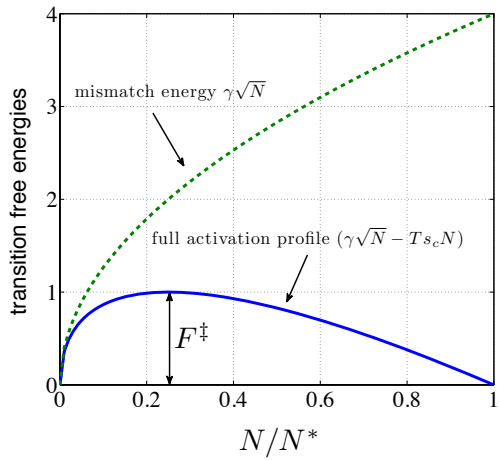}
        \vspace{-2mm}
        \caption{\label{FFN} \small Typical nucleation profile for
          structural reconfiguration in a supercooled liquid from
          Eq.~(\ref{FN}) and its surface energy component $\gamma
          \sqrt{N}$, normalized by the typical barrier height, which
          reaches $(35-37) k_B T$ at $T = T_g$.}
    \end{center}
\end{figure}

During an activated reconfiguration, two distinct low-energy
structures must coexist locally, implying that a higher free-energy
{\em interface} region must be present.  Because the interface
separates aperiodic arrangements, it has no obvious structural
signature; non-linear spectroscopy is generally required to detect the
interface.  The transitions proceed in a nucleation-like fashion; they
are driven by the multiplicity of the configurations, subject to the
mismatch penalty at the interface $\gamma\sqrt{N}$:\cite{KTW,
  LW_aging}
\begin{equation} \label{FN} F(N) = \gamma \sqrt{N} - T s_c N,
\end{equation}
see Fig.~\ref{FFN}.  Here, $N$ gives the number of beads contained
within the nucleus and the coefficient $\gamma = \frac{3}{2} \sqrt{3
  \pi} k_B T \ln(a^2/d_L^2\pi e)$.\cite{XW, LW} The lengths $a$ and
$d_L \simeq 0.1 a$ denote the volumetric size of a bead and the
Lindemann displacement\cite{L_Lindemann} respectively. The
reconfiguration time corresponding to the
the activation profile in Eq.~(\ref{FN}):
\begin{equation} \label{tau} \tau = \tau_0 e^{F^\ddagger/k_B T} =
  e^{\gamma^2/4 T s_c},
\end{equation}
grows in a super-Arrhenius fashion with lowering the temperature
because of the rapid decrease of the configurational
entropy\cite{PhysRevLett.64.1549, RichertAngell}
\begin{equation} \label{sc}
s_c \simeq \Delta c_p T_g(1/T_K - 1/T),
\end{equation}
where $\Delta c_p = T \prtl s_c/\prtl T|_{T_g}$ is the heat capacity
jump at $T_g$, per bead.  The quantity $T_K$, often called the
Kauzmann temperature or the ``ideal glass'' transition temperature,
corresponds to the temperature at which the configurational entropy of
an equilibrated liquid would presumably vanish: Hereby the log number
of alternative configurations would scale sub-linearly with the system
size. While the temperature $T_K$ appears in certain meanfield
models,\cite{MCT1} the ideal-glass state, if any,\cite{SWultimateFate}
would be impossible to reach in actual liquids because of the
diverging relaxation times, see Eq.~(\ref{tau}).

The nucleus size 
\begin{equation} \label{N*}
N^* \equiv (\gamma/Ts_c)^2, 
\end{equation}
where $F(N^*) = 0$, is special because it corresponds to a region size
at which the system is guaranteed to find at least one typical liquid
state. As a result, the size $N^*$ corresponds to the smallest region
that can reconfigure in equilibrium. Each transition requires work
$\gamma \sqrt{N}$ to create and grow the interface, at the expense of
relaxing the old interfaces. Thereby, the total number of interfaces
remains constant on average.  One thus concludes that the liquid
harbors one interfacial region with an associated excess free energy
$\gamma \sqrt{N^*}$ per region of size $N^*$, consistent with the
quench being a higher free energy state than the corresponding
crystal.  In fact, the total mismatch penalty in a sample of size $N$
is $\gamma \sqrt{N^*} (N/N^*) = T s_c N$ and thus equals the enthalpy
that would be released if the fluid crystallized at this temperature,
save a small ambiguity stemming from possible differences in the
vibrational entropies.  The cooperativity size $N^*$ grows with the
decreasing configurational entropy\cite{KTW} (see Eqs.~(\ref{sc}) and
(\ref{N*}): $N^*(T) \simeq N^*(T_g)[(T_g-T_K)/(T-T_K)]^2$. Still, it
reaches only a modest value of 200 or so at the glass transition on
one hour time scale.\cite{XW, LW} $N^*(T_g) = 200$ corresponds to a
physical size $\xi \equiv a (N^*)^{1/3} \simeq 6.0 \, a$, i.e. almost
universally about two-three nanometers. This important prediction of
the RFOT theory has been confirmed by a number of distinct
experimental techniques \cite{Spiess, RusselIsraeloff, CiceroneEdiger}
and recently, by \emph{direct} imaging of the cooperative
rearrangements on the surface of a metallic glass.\cite{Gruebele2010}
Now, the resulting concentration of the domain walls near the glass
transition is, approximately,
\begin{equation} \label{nDW} n_\text{DW}(T_g) \simeq 1/\xi(T_g)^3
  \simeq 10^{20} \mbox{cm}^{-3}.
\end{equation}

In the rest of the article we will assume the material is just above
its glass transition temperature $T_g$, so that it represents a very
slow, but equilibrated liquid. Below $T_g$, the lattice remains
essentially what it was at $T_g$, save some subtle changes stemming
from the decreased vibrational amplitudes and aging. 

The present authors have argued\cite{ZL_JCP} that the interfaces may
host special midgap electronic states, subject to a number of
conditions. These conditions are satisfied in many amorphous
chalcogenides and pnictides.\cite{ZLMicro1} The number of the
intrinsic midgap states is about 2 per interface, implying a
concentration of about $2/\xi^3$, where $\xi$ is the cooperativity
length from Eq.~(\ref{nDW}). The argument in Ref.\cite{ZL_JCP} is
independent from the present considerations and is based on the Random
First Order Transition (RFOT) theory of the glass
transition.\cite{LW_ARPC} These intrinsic states are centered on
under- or over-coordinated atoms and are surprisingly extended for
such deep midgap states. In addition, the states exhibit the reverse
charge-spin relation. The existence of the intrinsic states allows one
to rationalize a number of optoelectronic anomalies in chalcogenides
in a unified fashion.\cite{ZL_JCP} The interface-based electronic
states in glasses are quasi-one dimensional and are relatively
extended along that dimension.\cite{ZL_JCP} This quasi-one
dimensionality stems from the structural reconfigurations themselves
being quasi-one dimensional: Activated transitions between typical
configurations occur by a nearly unique sequence of single-atom moves
that are nearby in space.  In the simplest possible Hamiltonian that
couples electronic motions to the atomic moves, by which a transition
between two metastable states takes place, only \emph{relative}
positions of the atoms are directly coupled to the electronic density
matrix:\cite{ZL_JCP}
\begin{equation} \label{Hel} \hspace{-1mm} \cHel = \sum_{n, s} \left[
    (-1)^n \epsilon_n c_{n,s}^{\dagger}c_{n,s}^{\phantom{\dagger}} -
    t_{n, n+1} (c_{n,s}^{\dagger}c_{n+1,s}^{\phantom{\dagger}} +
    H.C. ) \right], \hspace{-2mm}
\end{equation}
where $c_{n,s}^{\dagger}$ ($c_{n,s}^{\phantom{\dagger}}$) creates
(removes) an electron on site $n$ with spin $s$.  The on-site energy,
which reflects local electronegativity, is denoted with $(-1)^n
\epsilon_n$. The electron transfer integral $t_{n, n+1}$ between sites
$n$ and $n+1$ depends on the inter-site distance $d_{n, n+1}$. The
sites are centered on beads, \emph{not} atoms. The beads are numbered
with index $n$, $1 \le n \le N^*$, in the order they would move during
the reconfiguration.

\section{Classification of coordination defects}
\label{parent}

The structural model proposed by us in Ref.\cite{ZLMicro1} presents a
systematic way to classify amorphous structures and the associated
electronic-structure peculiarities in vitreous
$pp\sigma$-networks. Without claiming complete generality, we will
consider the following specific type of parent structures, which are
largely based on Burdett and MacLarnan's model of (the crystalline)
black phosphorous and rhombohedral arsenic.\cite{burdett5764} In this
model, each vertex on a simple cubic structure is connected to exactly
three nearest neighbors, where only right angles are permitted between
the links. Two specific \emph{periodic} ways to place the links
according to this prescription correspond to the crystals of black
phosphorous and rhombohedral arsenic. We have seen that this model has
a significantly broader applicability, if one allows for distinct
atoms and also vacancies in the original cubic lattice.  For instance,
let us take the parent structure for black phosphorus (Fig.~5(b) of
Ref.\cite{ZLMicro1}), place pnictogens and chalcogens in the rock salt
fashion (as in Fig.~6(b) or 7 of Ref.\cite{ZLMicro1}), and then omit
every third pnictogen in a particular fashion (as in Fig.~7 of
Ref.\cite{ZLMicro1}).  This procedure yields \emph{both} the crystal
structure and the stoichiometry of the archetypal
pnictogen-chalcogenide compound Pn$_2$Ch$_3$ (``Pn''= pnictogen,
``Ch''=chalcogen). In the crystal, each pnictogen and chalcogen are
three- and two-coordinated, respectively, thus conforming to the octet
rule.\cite{Burdett1995} The actual structure of a crystal or glass is
thus viewed as a result of the following multi-stage
procedure:\cite{ZLMicro1} (1) start with the simple cubic structure
with all vertices linked to all six nearest neighbors; (2) break the
links and place vacancies to satisfy the stoichiometry and the octet
rule. The resulting, lowered-symmetry structure is called the
``parent'' structure. To generate the actual structure from the parent
structure, (3) shift the atoms slightly toward the linked vertices;
(4) turn on the interactions by associating an electronic transfer
integral to each bond; and (5) geometrically optimize the structure,
subject to the repulsion between the ionic cores and the variations in
the electronegativity, if any.

In the above prescription, it is generally impossible to satisfy the
stoichiometry and the octet rule at the same time except in the case
of periodic crystals, as we will see explicitly in a moment. As a
result, the resulting \emph{parent} structure will generally exhibit a
variety of defects in the form of under- and over-coordinated atoms,
such as three-coordinated chalcogens. Similarly, incorporating
elements of group 14, such as germanium, into the lattice of strictly
three-coordinated vertices would introduce coordination defects. The
just noted presence of four-coordinated vertices presents a convenient
opportunity to point out that the hereby proposed structural model,
though not unique, is special in the context of $pp\sigma$-bonded
solids in distorted octahedral geometry. Suppose that, instead, we
started from a model where each vertex is \emph{four}-coordinated. The
corresponding parent structures, if periodic, could be unstable toward
a \emph{tetrahedrally} bonded solid with the $\beta$-tin or related
structure.\cite{BurdettLeePeierls}
As a result, the present discussion is likely limited to compounds
where elements from group 14 are in the minority, except when these
elements come in pairs with chalcogens, such as in GeSe or
Ge$_2$Sb$_2$Te$_5$.  Such pairs are isoelectronic with a pair of
elemental pnictogens, and so both constituents of the pair are
three-coordinated. This is not unlike, for instance, how GaAs forms
the diamond (zincblende) structure, in which each atom is four
coordinated.

The three-coordinated parent structures on a simple cubic lattice,
with right angles between the bonds, have another very special
property:\cite{burdett5764} In any lattice satisfying this constraint,
whether periodic or not, each bond-containing line is a strictly
alternating bond/gap pattern. Informally speaking, one may think of
the lattice as made of linear chains, \emph{each} of which consists of
white and black segments of equal length in strict alternation; the
junctions between adjacent segments correspond to the lattice
vertices, while the white and black segments correspond to no-link
and link respectively. This observation allows one to easily estimate
the number of all possible distinct parent structures. Since there are
only two distinct ways to draw a perfect bond/gap alternating pattern,
the total number of the parent structures in a sample of size $L_1
\times L_2 \times L_3$ with no vacancies is $2^{(L_1 L_2+ L_1 L_3 +
  L_2 L_3)/a^2}$, where $a$ is the lattice spacing. Note that in
contrast to Ref.\cite{burdett5764}, we count as inequivalent two
structures that can be obtained from each other by rotation, because
they will both contribute to the phase space.  We immediately observe
that the total number of distinct parent structures is
sub-thermodynamic in that it scales exponentially with the
\emph{surface} of the sample, not its volume. As a result, such
defect-free structures can not contribute significantly to the library
of \emph{bulk} aperiodic states. The only exception to this statement
is the hypothetical Kauzmann, or ideal-glass state, in which the
configurational entropy would strictly vanish.\cite{Kauzmann} One thus
concludes that an equilibrium liquid must contain not only perfectly
bond-alternating configurations, but a finite number of defected
configurations per unit length in each line, where a ``defect''
consists of two or more bonds (gaps) in a row.

Furthermore, parent structures of certain common stoichiometries, such
as Pn$_2$Ch$_3$, must host a thermodynamic number of
vacancies.\cite{ZLMicro1} Adding vacancies to defectless
\emph{aperiodic} parent structures should generally lead to the
creation of coordination defects.  Indeed, to maintain the
stoichiometry, the vacancies should be spaced on average by the same
distance. On the other hand, one can construct a bond-breaking pattern
with an arbitrarily large period, which can be made arbitrarily
different from the average spacing between the vacancies.

\begin{figure}[t]
  \centering
\includegraphics[width=0.6 \figurewidth]{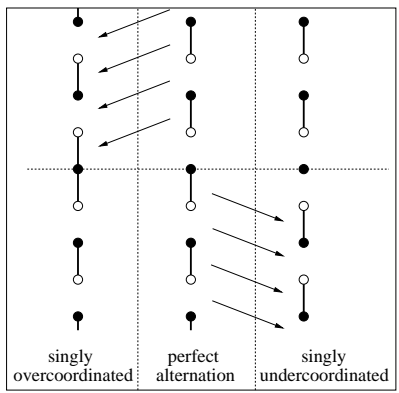}
\caption{Generation of singly malcoordinated atoms by a
  ``cut-and-shift'' procedure, see text. The filled and empty circles
  do not necessarily imply chemically distinct species, but are used
  to emphasize that it is the bond pattern that is shifted, not the
  atoms themselves.}
\label{shift}
\end{figure}

As a consequence of the deviation from the strict alternation pattern
along the individual linear chains, a thermodynamic number of atoms in
a representative parent structure must be either under- or
over-coordinated. Note that after geometric optimization, defects in
the parent structure will be generally mitigated or even removed. For
example, at least one of the covalent bonds around a four-coordinated
pnictogen will generally elongate in the deformed structure, in the
pnictogen's attempt to attain the favorable valency 3. As a result,
the overcoordination will be spread among a larger number of atoms. An
example of defect annihilation is when an extra bond on an
overcoordinated atom annihilates with a missing bond on a nearby
undercoordinate vertex.  For these reasons, a defect in the deformed
lattice is not well-defined. Here, we analyze several important defect
configurations in the \emph{parent} structure, whereby the local
coordination is entirely unambiguous. Before proceeding, we should
note that the discussion is not limited to atoms of valence three and
two. The above discussion applies even if a fraction of the vertices
in the defectless parent structure shoud be four-coordinated, as would
be appropriate for elements of group 14. Estimating the total number
of distinct parent structures is no longer straightforward, however it
is clear the number could be only lowered compared to the strict
Burdett-McLarnan case.

\begin{figure}[t]
\centering
\includegraphics[width= 0.6 \figurewidth]{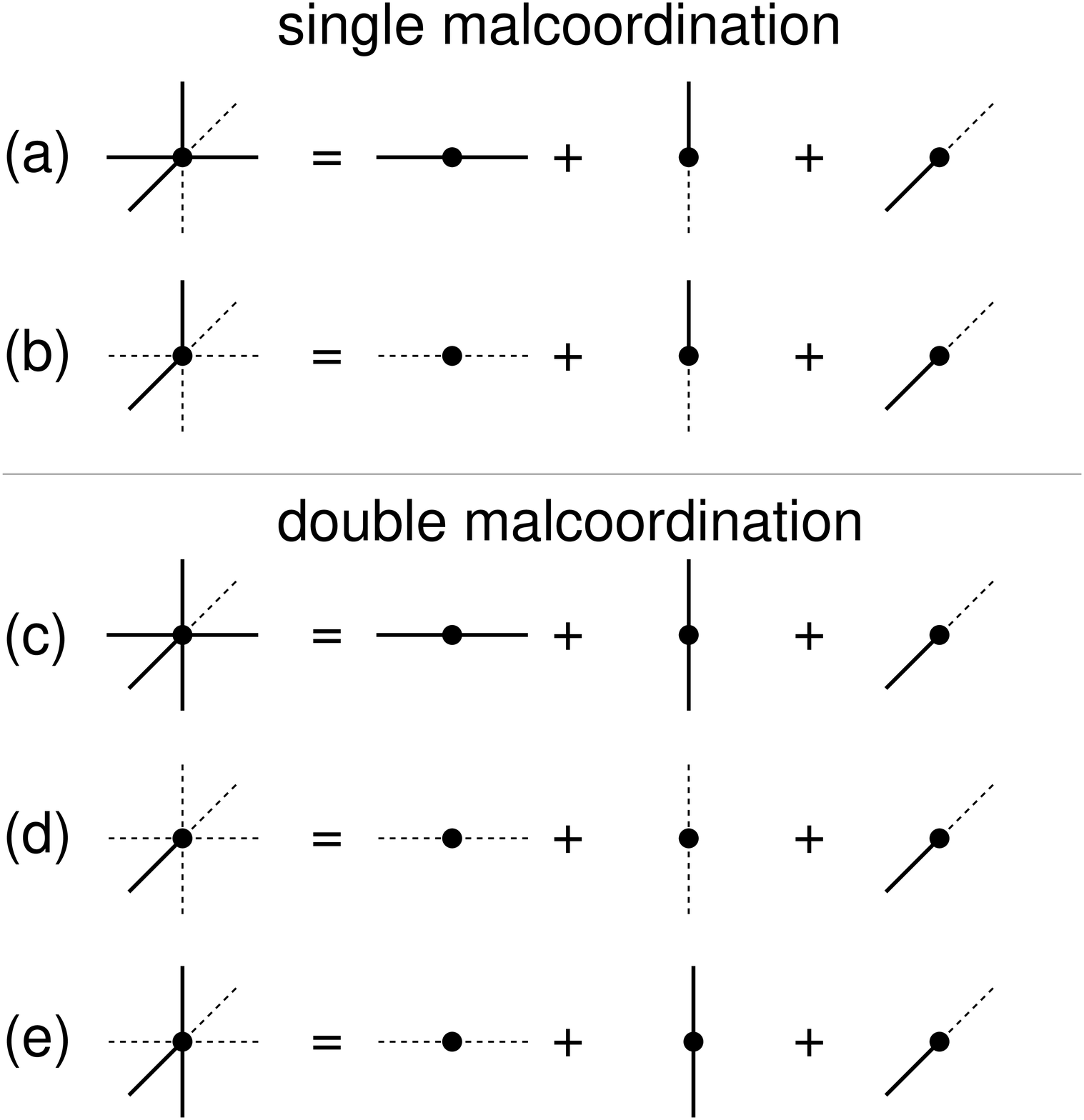}
\caption{Examples of malcoordination defects of different orders on a
  pnictogen site and their relation to single-malcoordination defects
  along individual directions. A similar diagram can be drawn for
  chalcogens, which are doubly coordinated in a defectless parent
  structure, the bond angle equal to $90^\circ$.}
\label{CoordVariants}
\end{figure}

Let us begin the discussion of defected parent structures with an
ideal parent lattice with one singly overcoordinated vertex. This
defect corresponds to two links in a row in one of the three chains
crossing the vertex in question. Formally, one can obtain such a
defect from a perfect Burdett-MacLarnan structure, for example, by
choosing a vertex, removing the adjacent gap-segment from one of the
chains and shifting the rest of the bond pattern on that side toward
the vertex, see Fig.~\ref{shift}. Note that owing to
$pp\pi$-interactions, such shift generally invokes an energy cost or
gain in \emph{addition} to the cost of overcoordinating the chosen
vertex, which we will discuss later.  One may obtain a singly
undercoordinated vertex analogously, see Fig.~\ref{shift}.  It is
straightforward to see that more complicated, multiple-malcoordination
configurations can always be presented as superpositions of the
simple, single-malcoordination configurations described above, see
Fig.~\ref{CoordVariants}. For instance, a doubly-overcoordinated atom
simply corresponds to overcoordination on two intersecting chains, at
the vertex in question.  Another common type of defect could be
obtained by simply replacing a gap in an ideal parent structure by a
link. This defect is equivalent to \emph{two} singly-overcoordinated
atoms along the same line next two each other. Coordination defects of
higher yet order can be constructed similarly. Note that bond angles
other than $90^\circ$ also amount to a coordination defect, even if
the total number of bonds obeys the octet rule, see
Fig.~\ref{CoordVariants}(e).

\begin{figure}[t]
\centering
\includegraphics[width= 0.8 \figurewidth]{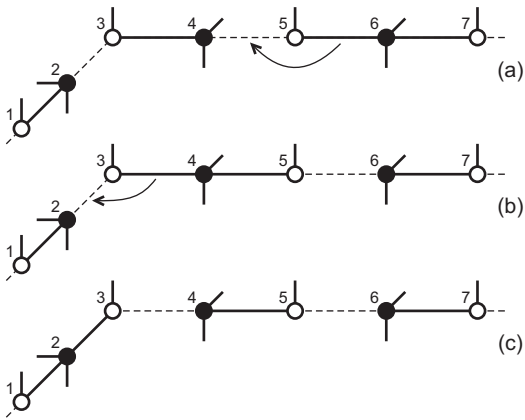}
\caption{Illustration of the motion of an overcoordination defect by
  bond-switching along a linear chain (from atom 6 in (a) to atom 4 in
  (b)) or making a turn (from atom 4 in (b) to atom 2 in (c)).}
\label{bondswitching}
\end{figure}

\begin{figure}[t]
\centering
\includegraphics[width=0.4\figurewidth]{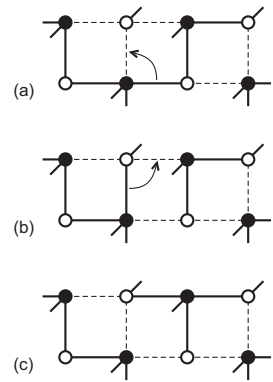}
\caption{A specific example of the motion of a topologically stable
  pair of an over- and under-coordinated defect, whereby the two
  defects move to the right while switching chains.  }
\label{topstable}
\end{figure}

Single-malcoordination configurations can move along linear chains or
make turns by bond switching, see Fig.~\ref{bondswitching}, whereby
the defect moves by two bond lengths at a time.  Direct inspection
shows, however, that an attempt to make a turn at a pnictogen in
Fig.~\ref{bondswitching} (such as atom 4) will create a double defect,
such as in Fig.~\ref{CoordVariants}(e), and is therefore likely to be
energetically unfavorable. Generally, only turns on atoms of the same
parity as the defect center are allowed. Note that the distance
between the nearest singly over- and under-coordinated atoms, along a
chain, should be always an even number of bonds, if the
malcoordination is along the very same chain. These ``opposite''
defects can be brought together and mutually annihilated. Conversely,
one observes that two opposite defects separated by an odd number of
bonds can not annihilate. A pair of such defects can be said to be
topologically stable, see Fig.  \ref{topstable}. In this Figure, we
also illustrate that this pair of defects is mobile, subject to the
``turning'' rule above, of course.  In addition, the specific
defect-pair configuration in Fig.~\ref{topstable} is special because
it allows for defect motion without the creation of $\Pi$-shaped bond
motifs, in which three bonds constitute three sides of a rectangle.
These motifs are somewhat energetically unfavorable, see the Appendix.
Finally note that one of the nearest neighbors of the four-coordinated
atom in Fig.~\ref{topstable}(c) is not covalently bonded to any other
atoms, implying the orientation of its bond with the central atom is
relatively flexible. We will see below that owing to this extra
flexibility and the said topological stability of the defect pair, the
central atom is particularly unstable toward $sp^3$ hybridization when
positively charged. Nevertheless, since each of the individual defects
can be removed, such $sp^3$ hybridized units do not represent a
distinct type of defect. Conversely, the formation of $sp^3$
hybridized atoms in $pp\sigma$-networks is \emph{reversible}.

We thus conclude that singly-malcoordinated atoms comprise the
\emph{only} distinct type of defect in the parent structure.  The
number of the motifs in the actual deformed structure that originate
from the coordination defects in the parent structure is determined by
the corresponding free energy cost. This cost consists of the energy
cost proper of single-malcoordination defects, the energy of their
interaction, and a favorable entropic contribution reflecting the
multiplicity of distinct defect configurations. With regard to the
interaction, it is easy to see that like defects should repel and
defects of the opposite kind should attract. Indeed, the higher the
over (under) coordination is, the larger the energy cost; conversely,
mutual annihilation of a singly over- and under-coordinate defects is
energetically favorable. One may thus conclude that for both energetic
and entropic reasons, there are no bound states of
single-malcoordination defects to speak of that could be classified as
a distinct type of defect. The only such bound state is a
topologically-stable pair of a singly over- and under-coordinated
vertex, as in Fig.~\ref{topstable}.

We have pointed out earlier\cite{ZLMicro1} that defectless parent
structures can be formally represented as a subset of configurations
of the 64-vertex model, which is the 3D generalization of the 6-vertex
model of ice and 8-vertex model of anti-ferroelectrics.\cite{Baxter}
The presence of malcoordination can still be implemented in the
64-vertex model, by assigning an appropriate energy cost to such
coordination.  However, the presence of $pp\pi$-interaction implies
non-adjacent bonds interact directly. One concludes that parent
structures of pertinence to $pp\sigma$-networks are subject to a more
general class of models, such as the Ising model with next nearest
interactions.\cite{Baxter}

In concluding the discussion of defect classification, we emphasize
that stoichiometry-based vacancies in parent structures should be
regarded not as defects, but as an intrinsic part of the parent
structure. As in the example of the parent Pn$_2$Ch$_3$
structure,\cite{ZLMicro1} such stoichiometry-based vacancies
contribute to driving the distortion of the original simple cubic
lattice and are minimized in the deformed structure, subject to
competing interactions. Formally, this means the energy cost of local
negative density fluctuations in the deformed structure that stem from
the vacancies in the parent structure is a \emph{perturbation} to the
$pp\sigma$ bonding, similarly to the $sp$-mixing.  Elementary
estimates of the energy cost of a vacancy in a
glass\cite{LWEchoComment} show the cost is prohibitively high,
dictating that an equilibrium melt above the glass transition will
contain negligibly few voids of atomic size or larger. Similarly,
homopolar bonds are not regarded as defects from the viewpoint of the
present structural model. In fact, we have argued\cite{ZLMicro1} a
thermodynamic quantity of homopolar bonds in parent structures are
required for the presence of transport in supercooled melts. These
notions are consistent with the relatively low electronegativity
variations in $pp\sigma$-bonded materials\cite{ZLMicro1} and hence a
weak applicability of Pauling's rules.\cite{Burdett1995} To avoid
possible confusion, we reiterate that the present argument strictly
applies only to equilibrium melts or to glass obtained by quenching
such melts. Amorphous films made by deposition generally do not
correspond to an equilibrated structure at any temperature and,
presumably, could host a greater variety of structural motifs.

\section{The intrinsic midgap electronic states}
\label{soliton}

\begin{figure}[t]
\centering
\includegraphics[width=\figurewidth]{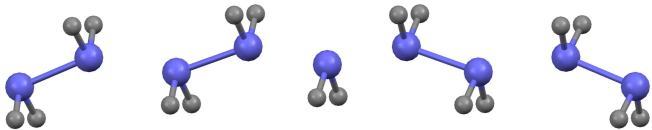}
\caption{Central part of neutral (AsH$_2$)$_{21}$ chain, whose ground
  state contain a coordination defect and the associated midgap
  level. The calculations are performed on semiempirical level by
  MOPAC\cite{MOPAC} with PM6 parametrization and the RHF/ROHF method,
  see Ref.\cite{ZLMicro1} for full computational detail and
  Supplementary Material \cite{SupplMat} for comparison between the
  results of optimization for several chain lengths.}
\label{fig_AsH2chain}
\end{figure}

To characterize the motifs arising in the deformed structure as a
result of one singly undercoordinated atom in a parent structure, we
consider first an \emph{isolated} chain of $pp\sigma$-bonded arsenic
atoms passivated by hydrogens: (AsH$_2$)$_n$. For concreteness, the
chain is oriented along the $z$ axis. An infinite undistorted chain is
Peierls-unstable.\cite{ZLMicro1} The two ground state configurations
of an infinite (AsH$_2$)$_n$ chain consist each of a perfect
alternation pattern of covalent and secondary bonds.  A good approximation for
these ground states can be obtained using a finite linear chain
containing an even number of monomers, see the preceding
article.\cite{ZLMicro1}

Even if the chain is embedded in a lattice, each of these two ground
states can still be regarded as the result of a Peierls transition, so
long as the deviation from the strict octahedral coordination is weak.
It is shown in Ref.\cite{ZLMicro1} that the $pp\sigma$-chain portion
of the full one-particle wave-function is an eigenfunction of the
effective Hamiltonian:
\begin{equation} \label{renorm} \widetilde{\cH}=\cH_{pp\sigma} +
  \cV^\dagger \left( E - \cH_\text{env} \right)^{-1} \cV
\end{equation}
with the same eigenvalue $E$, where $\cH_{pp\sigma}$ contains
exclusively the $p$ orbitals in question, $\cH_\text{env}$ the rest of
the orbitals, and $ \cV$ the transfer integrals between the two sets
of orbitals. Using Eq.~(\ref{renorm}), the effect of the environment
can be presented as an (energy-dependent) renormalization of the
on-site energies and transfer integrals of the $pp\sigma$-network
proper, see Appendix C of Ref.\cite{ZLMicro1} The most significant
contributor to this perturbation is the competing $sp$ interactions.
The perturbation resulting from these competing interactions turns out
to be sufficiently weak, as could be inferred from (a) its magnitude
being close to the corresponding estimate using a perturbation series
in the lowest non-trivial order; and (b) the bond morphology in the
chain conforming to what is expected from a
$pp\sigma$-network.\cite{ZLMicro1}

\begin{figure}[t]
\centering
\includegraphics[width=0.9 \figurewidth]{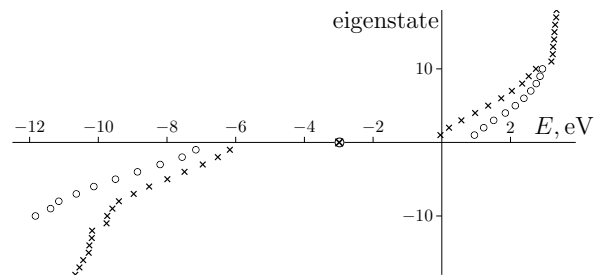}
\caption{Electronic energy levels of the defected chain
  (AsH$_2$)$_{21}$ chain from Fig.~\ref{fig_AsH2chain}: full MO
  calculation (crosses) vs.  one-orbital model with renormalized
  $pp\sigma$ integrals (circles). States below the gap are filled; the
  midgap state is half-filled.  }
\label{fig_AsH2ev}
\end{figure}

\begin{figure}[t]
\centering
\includegraphics[width=0.9 \figurewidth]{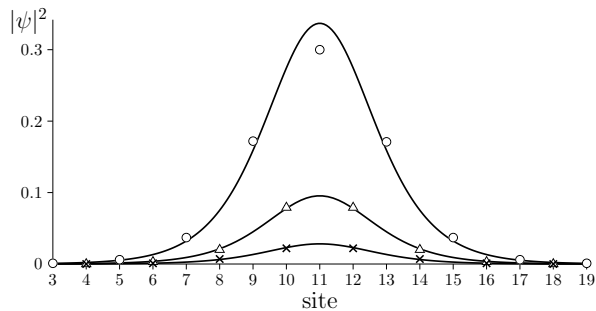}
\caption{The wave function squared of the midgap state of the
  (AsH$_2$)$_{21}$ chain: the circles correspond to the arsenics'
  $p_z$ atomic orbitals (AO) (total contribution 73\%), triangles As
  $s$-AO's (21\%), crosses -- the rest of AO's (6\%). All
  contributions are nearly zero on every other atom; these very small
  values are omitted for clarity. The solid line corresponds to
  Eq.~(\ref{psishape}).}
\label{elwavefunction}
\end{figure}

Now, in contrast to an even-numbered chain, the ground state of a
$(4n+1)$-membered open chain can be thought of as having one singly
undercoordinated atom, see Fig.~\ref{fig_AsH2chain}, while the ground
state of a $(4n+3)$-membered open chain contains one singly
overcoordinated atom.  Either of these ``defected'' chains is a
neutral radical with an unpaired spin.  The spectrum of the defected
chain, shown with crosses in Fig.~\ref{fig_AsH2ev}, exhibits a
singly-populated state exactly in the middle of the forbidden gap. The
electronic wave function of this midgap state, shown in
Fig.~\ref{elwavefunction}, is centered on the undercoordinated atom
and exhibits a significant degree of delocalization.  Likewise, the
deviation from the perfect bond alternation, although the strongest in
the middle of the chain, is delocalized over several atoms. The
partitioning of the electronic density between the $p_z$ orbitals
proper (73\%), the arsenics' $s$-orbitals (21\%), and the rest of the
orbitals (6\%) indicate that the interactions that compete with the
$pp\sigma$-bonding proper are significant but, nevertheless, may be
regarded as a perturbation to the latter bonding, similarly to the
perfectly dimerized chain.\cite{ZLMicro1} Note that the position of
the nodes of the wave-function is easy to understand using elementary
H\"{u}ckel considerations: Consider an odd-numbered chain of identical
atoms, each hosting an odd number of orbitals. A moment thought shows
that the resulting set of orbitals (in one electron approximation) has
a half-filled non-bonding orbital, whereby the wave-function has nodes
on every second atom.

\begin{figure}
\includegraphics[width= 0.8 \figurewidth]{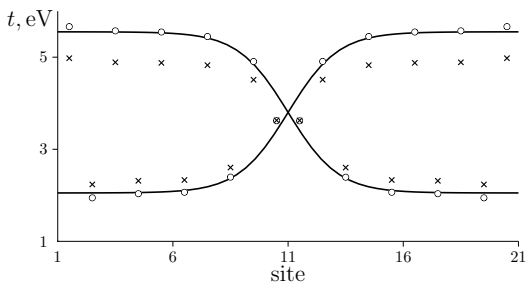}
\caption{Spatial dependence of the $pp\sigma$ transfer integral in the
  (AsH$_2$)$_{21}$ chain. The actual transfer integrals from
  $\cH_{pp\sigma}$ in Eq.~\ref{renorm} are shown with crosses. The
  renormalized integrals from $\widetilde{\cH}$ in Eq.~\ref{renorm} are
  shown with circles. The solid line corresponds to
  Eq.~(\ref{kinkshape}). }
\label{fig_AsH2kink}
\end{figure}

The delocalization of the malcoordination is manifest in the broad
sigmoidal profile of the $pp\sigma$ transfer integral as a function of
the coordinate, see Fig.~\ref{fig_AsH2kink}.  Because the
malcoordination is distributed among a large number of bonds in the
deformed structure, the resulting structural signature of such a
defect is far from obvious.  In fact, the magnitude of the transfer
integrals for the bonds adjacent to the overcoordinated atom in a $(4n
+3)$ chain is very similar to that for the undercoordinated atom in a
$(4n + 1)$ chain. This magnitude is approximately the average of the
magnitude of the transfer integral for the covalent and secondary bond
in a perfectly dimerized chain, see Fig.~\ref{fig_AsH2kink}. The same
comment applies to the corresponding bond lengths in the first place,
see Fig.~\ref{widths12}, in view of the direct relationship between
$d$ and $t$.  As a result, the malcoordination is essentially fully
``absorbed'' by the chain. For instance, the distances between atoms 1
and 19 in 19- and 21-member chains are 48.71 and 48.70 \AA
~respectively. These notions emphasize yet another time the ambiguity
of defining coordination or coordination defects in actual, deformed
structures.
Despite being relatively extended, the undercoordination defect could
be thought of as separating two alternative states with perfect
dimerization. This defect exhibits the reverse charge-spin relation,
because it is neutral when its spin is equal 1/2. Adding or removing
an electron would yield a singly-charged defect with spin 0. These
characteristics of the midgap state in the (AsH$_2$)$_n$ chain are
entirely analogous to those of the midgap state in
trans-polyacetylene, which is a classic example of a Peierls
insulator.  Until further notice, we will consider only neutral
chains. Unlike the perfectly dimerized chain, a geometry optimized
defected chain is somewhat curved on average; this effect is small,
however.

The surprisingly large spatial extent of the midgap state in the
arsenic chain results from an interplay of several factors, similarly
to how the spatial characteristics of the solitonic state in
polyacetylene depends on the lattice stiffness, electronic
interactions etc. Nevertheless, it turns out that in both systems the
extent of delocalization is determined essentially by only two
parameters, namely the renormalized transfer integrals of the covalent
and secondary bond in the perfectly dimerized chain, denoted with
$\rt$ and $\rt'$ respectively. (In the case of polyacetylene, the
transfer integrals are $pp\pi$, of course.)  This simplification takes
place because, similarly to the case of a perfectly dimerized chain,
the effects of the competing intra-chain and external interactions on
the midgap state largely reduce to the renormalization of the
$pp\sigma$ interaction proper.  To demonstrate this notion, we take
the geometrically-optimized chain, extract only the $pp\sigma$
integrals, and renormalize them according to Eq.~(\ref{renorm}) while
setting $E$ equal exactly to the center of the forbidden gap. The
spectrum of the resulting Hamiltonian is shown with circles in
Fig.~\ref{fig_AsH2ev}. Setting $E$ at the center of the gap should
result in an error the greater, the further the state in question from
the gap center. Yet one infers from Fig.~\ref{fig_AsH2ev} that the
error in the spectrum is relatively small, consistent with the
smallness of the perturbation.

The spatial dependence of the renormalized $pp\sigma$ integrals is
shown in Fig.~\ref{fig_AsH2kink} with circles.  In the continuum
limit\cite{PhysRevB.21.2388} of the Su-Schreiffer-Heeger (SSH) model
for midgap states in trans-polyacetylene,\cite{PhysRevB.21.2388} the
transfer integral depends on the coordinate according to the
expression\cite{ZL_JCP}:
\begin{equation} \label{kinkshape}
  \rt_n=\frac{\rt+\rt'}{2}+\frac{\rt-\rt'}{2}(-1)^n
  \tanh\left(\frac{n-n_0}{\xi_s/d}\right),
\end{equation}
where the soliton is centered on bond $n_0$ and its half-width $\xi_s$
is given by:\cite{ZL_JCP}
\begin{equation} \label{xi_SSH}
  \xi_s=\frac{\rt+\rt'}{\rt-\rt'} d,
\end{equation}
where $d$ is the length of the bond in the parent structure.  Using
the renormalized values of the parameters $\rt$ and $\rt'$ from
Ref.\cite{ZLMicro1}, we obtain the solitonic profile shown by the
solid line in Fig.~\ref{fig_AsH2kink}.  Note the difference between
the two sets of transfer integrals is small.  Finally, the electronic
wavefunction in the same continuum limit is given by the function:
\begin{equation} \label{psishape} \psi_n^2 = (1/2) (d/\xi_s)
  \cosh^{-2}[ (n-n_0)d/\xi_s],
\end{equation}
as shown with the solid lines in Fig.~\ref{elwavefunction}. Again, the
agreement with the result of an electronic structure calculation using
renormalized transfer integrals is good. The agreement is not expected
to be perfect, however, because of the approximations inherent in the
continuum limit\cite{PhysRevB.21.2388} of the SSH model.

We thus observe that to describe both the ground state and defected
configurations of linear chains from a $pp\sigma$ network, we can
apply a Hamiltonian that contains explicitly only the $pp\sigma$
interactions whose parameters are renormalized by other interactions.
This resulting tight-binding Hamiltonian, which corresponds with the
renormalized Hamiltonian (\ref{renorm}), reads as:
\begin{equation} \label{Helatom} \cHel = \sum_{n, s} \left[ (-1)^n
    \epsilon_n c_{n,s}^{\dagger}c_{n,s}^{\phantom{\dagger}} - \rt_{n,
      n+1} (c_{n,s}^{\dagger}c_{n+1,s}^{\phantom{\dagger}} + H. C. )
  \right],
\end{equation}
where $c_{n,s}^{\dagger}$ ($c_{n,s}^{\phantom{\dagger}}$) creates
(annihilates) an electron with spin $s$ in the $p_z$-orbital on atom
$n$. The on-site energy, which reflects the local electronegativity,
is denoted with $(-1)^n \epsilon_n$.  The renormalized electron
transfer integral $\rt_{n, n+1}$ between sites $n$ and $n+1$ depends
on the bond length $d_{n, n+1}$. The latter is determined by the
restoring force of the lattice, in addition to the Peierls symmetry
breaking force stemming from the electronic degree of freedom in
Eq.~(\ref{Helatom}). 
The restoring force of the lattice includes both the intra- and
extra-chain perturbations to the $pp\sigma$ interaction. We have seen
in Ref.\cite{ZLMicro1} that a portion of the intra-chain perturbation,
namely the $sp$-mixing also contributes to the symmetry-breaking that
results in bond-dimerization of the chain. Yet according to
Eqs. (\ref{xi_SSH}) and (\ref{psishape}), the participating competing
interactions determine the spatial characteristics of the coordination
defect largely through the values of only two parameters: $t$ and
$t'$, in addition to the lattice spacing $d$.

We are now ready to argue that the defect states centered on singly
malcoordinated atoms in $pp\sigma$-chains, as in
Figs.~\ref{fig_AsH2ev}-\ref{fig_AsH2kink} should be identified with
the intrinsic states proposed in Ref.\cite{ZL_JCP} On the one hand,
$pp\sigma$-bonded glasses meet the sufficient conditions for the
presence of the intrinsic states, as argued in the preceding
article.\cite{ZLMicro1} On the other hand, we have seen here a singly
malcoordinated atom is the \emph{only} type of defect present in
$pp\sigma$-bonded glasses, while all of its characteristics are
identical to those established independently for the intrinsic states,
including the reverse charge-spin relation and the relative
delocalization.  In addition, both types of states emerge at the
coexistence region of two distinct lowered-symmetry states, which
originated from a higher symmetry state.  In the rest of this Section,
we discuss the microscopic insights arising from both the common
aspects and complementarity of the semi-phenomenological approach from
Ref.\cite{ZL_JCP} (case 1) and the present, molecular model for the
midgap states in $pp\sigma$-bonded glasses (case 2).

Let us begin with the electronic Hamiltonian governing the formation
of the midgap states.  In both cases, the electronic Hamiltonians are
quasi-one dimensional, whereby the sites are situated along a deformed
line in space. Although the Hamiltonians are identical notation-wise,
they are distinct microscopically: In case 1, Eq.~(\ref{Hel}), the
sites refer to rigid molecular units, often called ``beads.'' By
construction, the beads are not significantly disturbed by liquid
motions and, conversely, interact only weakly with each other,
comparably to the van der Waals coupling. Beads usually contain
several atoms each; they are essentially chemically equivalent and
fill out the space uniformly. The volumetric size of the bead is
denoted with $a$.\cite{LW_soft} In contrast, in case 2,
Eq.~(\ref{Helatom}) the sites are actual atomic orbitals which are
generally not chemically equivalent.  To check whether the two
Hamiltonians are consistent, we recall the RFOT theory's prediction
that the smallest cooperativity size $\xi$ is just above
$2a$,\cite{LW_soft} implying a supercooled melt can host at most one
malcoordinated atom per $\sim 4$ beads at the onset of activated
transport at a temperature $T_\text{cr}$. The volumetric size $a$ of
the bead can be estimated based on the RFOT's prediction that the
configurational entropy at the glass transition on one hour time scale
is universally 0.8 per bead.\cite{XW} For As$_2$Se$_3$, this estimate
yields about a half of the As$_2$Se$_3$ unit per bead, i.e. roughly
one bead per arsenic atom, and $a \simeq 4.0$ \AA.\cite{ZL_JCP} An
inspection of the parent structure of As$_2$Se$_3$ in Fig. 8 of
Ref.\cite{ZLMicro1} demonstrates that it is possible to accommodate
one malcoordination defect per 4 arsenics so that the defects are
separated by at least two atoms. Since the defects are not immediately
adjacent in space, the geometric optimization of the parent structure
will not necessarily lead to their mutual annihilation.

Next, we address the width of the interface and the related
restriction on the spatial variation of the electronic transfer
integral. In Ref.\cite{ZL_JCP}, we used the simplest spatial profile
of bead displacement during a cooperative rearrangement to suggest
that the soliton half-width $\xi_s$ is bounded from below by $\xi/2$,
which is about $3a$ at $T_g$.  Inspection of Fig.~\ref{fig_AsH2kink}
indicates that the soliton, though still extended, is somewhat more
narrow, i.e. $\xi_s \simeq 2a$. (Recall also that $d < a$.) Using the
condition\cite{ZL_JCP}
\begin{equation} \xi_s < \frac{t+t'}{t-t'} a,
\end{equation}
we then should revise the lower limit on the variation of the transfer
matrix element to:
\begin{equation} \label{tratio} t'/t \gtrsim 0.3.
\end{equation}
We have seen earlier\cite{ZLMicro1} that the $pp\sigma$ bonded
materials in question do indeed meet this revised condition very well.
Finally we note that the prediction of Ref.\cite{ZL_JCP} that the
extent of the wave function somewhat exceeds $2\xi_s$ is indeed borne
out by the result in Fig.~\ref{elwavefunction}.

The RFOT theory makes specific predictions as to the free energy cost
of a structural reconfiguration. For instance, the total mismatch
penalty near the glass transition, according to the discussion of
Fig.~\ref{FFN}, is four times the typical barrier for the
reconfiguration. At the glass transition, this corresponds to about
$140 k_B T_g$, or generically about 3 to 4 eV. For the present
microscopic picture to be valid, this energy should exceed the energy
cost of the structural defect, which includes the energy of the
associated lattice deformation and the electronic energy proper.  We
have established earlier\cite{ZL_JCP} that the electronic energy
associated with the intrinsic state can be estimated via the optical
gap $\Eg$ and the effective attraction $\eU$ between like particles
occupying the state: $\sim (\Eg - |\eU|) \simeq 0.7 \Eg$. This figure
is about one quarter of the full interface energy predicted by the
RFOT theory. Alternatively, according to the solution of the continuum
limit of the SSH model, the cost of the intrinsic state on an isolated
defect can be expressed through its width $\xi_s$:
\begin{equation} \label{energyXI}
  \frac{4}{\pi}\frac{t}{1+\xi_s/a},
\end{equation}
yielding a similar figure to the above estimate.  One observes that,
indeed, the present microscopic picture is internally
consistent. There yet is another potential electronic contribution to
the defect energy.  We have already mentioned that the creation or
motion of a defect may by accompanied by a creation of $\Pi$-like
patterns of bonds in the parent structure; the latter are analyzed in
the Appendix.

\begin{figure}[t]
\centering
\includegraphics[width= 0.9 \figurewidth]{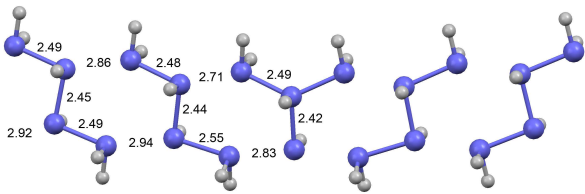}
\caption{Central portion of a geometrically optimized double chain of
  hydrogen-passivated arsenics. The top and bottom chains, containing
  19 and 17 arsenics, host an over- and under-coordinated arsenic
  respectively, c.f. Fig.~\ref{fig_AsH2chain}.  The numbers denote the
  bond lengths in Angstroms.}
\label{topstableAsH}
\end{figure}

A chain hosting a singly under- and over-coordinated defect can always
lower its energy via mutual annihilation of the defects, consistent
with conclusions of the continuum approach of Ref.\cite{ZL_JCP} Two
such defects on different chains, however, may be topologically stable
against such annihilation, as in Fig.~\ref{topstable}(c). Such a
configuration presents a felicitous opportunity to examine the
interaction of the defects in the \emph{deformed} structure, for
instance, by geometrically optimizing a double $(4n+1)-(4n+3)$ chain
of hydrogen-passivated arsenics, see Fig.~\ref{topstableAsH}. One
observes that in the ground state of a resulting defect pair, the
under- (over-) coordinated atom is negatively (positively) charged, so
as to obey the octet rule, see also Section \ref{sec.model}. The
resulting dipole moment of the chain is $3.3$ Debye; see the shapes of
the HOMO/LUMO orbitals and electrostatic map of the chain in the
Supplementary Material.\cite{SupplMat} The As-As angles at the
overcoordinated arsenic are 110, 110, and 128 degrees. Note in perfect
tetrahedral coordination, these angles would be 109.5 degrees. In
addition, the defects become more localized when in close proximity,
as shown in Fig.~\ref{widths12}, because of the partial cancellation
of the malcoordination. Alternatively, one may say the energy cost of
this localization partially offsets the stabilization resulting from
the binding of the defects.

\begin{figure}[t]
\centering
\includegraphics[width= 0.9 \figurewidth]{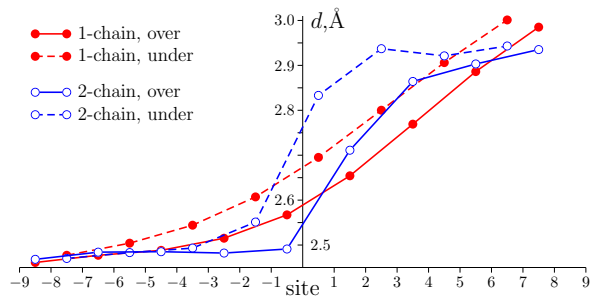}
\caption{Site dependence of the As-As bond length in single and double
  chains of hydrogen-passivated arsenics, as labeled by ``1-chain''
  and ``2-chain'' respectively. The ``1-chain, under'' corresponds to
  a (AsH$_2$)$_{17}$ chain similar to the (AsH$_2$)$_{21}$ chain from
  Fig.~\ref{fig_AsH2chain} and ``1-chain, over'' corresponds to a
  (AsH$_2$)$_{19}$ chain. The double-chain entries pertain to the
  individual chains in Fig.~\ref{topstableAsH}, ``over'' and ``under''
  correspond to the top and bottom chain respectively.}
\label{widths12}
\end{figure}

The extended nature of the coordination defects is consistent with
their mobility: In the continuum limit, the pinning of a soliton is in
fact strictly zero.  Consistent with the latter notions, the energy
gradients during geometrical optimization of the (AsH$_2$)$_n$ chain
above have very small numerical values. Notwithstanding this
complication, an estimate for the barrier for soliton hopping event
(by two bond lengths) can be made within the SSH model: Using the
effective transfer integrals from Eqs.~(\ref{kinkshape}) and
(\ref{xi_SSH}) we estimate the energy of the transition state
configuration (corresponding to a move by one bond length) at
approximately 0.1 meV, which is about $10^{-5} \Eg$. $pp\pi$
interactions may also contribute to pinning, see the Appendix for
quantitative estimates.  While the above notions may apply relatively
directly to neutral defects even in actual 3D lattices, the situation
appears to be more complicated for charged solitons (which in fact
greatly outnumber the neutral ones\cite{ZL_JCP}) because of lattice
polarization. This is work in progress.  Finally, note the high
mobility of the soliton seems particularly important in the context of
cryogenic anomalies of glasses, which have been argued to arise from
the low-barrier subset of the structural reconfigurations.\cite{LW,
  LW_RMP}

As an additional dividend of the present discussion, let us see that
the presence of an electronic - and hence $T$-independent -
contribution to the RFOT's mismatch penalty between distinct low free
energy aperiodic configurations allows one to partially explain the
deviation of chalcogenides from the RFOT-predicted correlation between
the thermodynamics and kinetics of the glass transition.  Indeed,
because of this $T$-independent contribution we expect corrections to
the detailed temperature dependence of the reconfiguration barrier
$F^\ddagger$ from Eq.~(\ref{tau}) and the value of the so called
fragility coefficient $m = (1/T)\prtl \log(\tau)/\prtl
(1/T)|_{T=T_g}$.\cite{LW_soft, StevensonW} The RFOT theory predicts
that this coefficient should be equal to what one may call a
``thermodynamically'' determined fragility $m_\sthermo = 34.7 \times
\Delta c_p(T_g)$ where $\Delta c_p(T_g)$ is the heat capacity jump at
the glass transition {\em per bead}.\cite{LW_soft, StevensonW} Using
the measured $\Delta c_p(T_g)$ {\em per mole}, and the bead size
determination from the fusion enthalpy produces excellent agreement
for dozens of substances,\cite{StevensonW} however among the outliers
are the chalcogenides. For instance, for As$_2$Se$_3$, the measured $m
\simeq 40$, while the theoretically computed $m_\sthermo \simeq
7.5$. Determination of the bead size is ambiguous in As$_2$Se$_3$,
because the corresponding crystal is highly anisotropic. In fact, in
the fusion-entropy based estimate, the bead comes out to be a single
atom, in conflict with the aforementioned requirement of chemical
homogeneity and with Ref.\cite{BL_6Spin}, where a bead was argued to
contain at least two atoms.  Using the RFOT's prediction that
$s_c(T_g) \simeq 0.8 k_B$ per bead, gives a better value $m_\sthermo
\simeq 17$,\cite{ZL_JCP} still not enough to account for the
disagreement.  Simple algebra shows that in the presence of a
$T$-independent contribution to the surface tension, the estimate of
the thermodynamic fragility from Ref.\cite{LW_soft} should be
increased by a factor of $[1 + 2 (\delta \Sigma) (T_g - T_K)/T_K]$,
where $\delta \Sigma$ is the relative contribution of the electronic
states to the total surface energy at $T_g$. Assuming the mentioned
$\delta \Sigma \simeq 0.25$ and using parameters for As$_2$Se$_3$ from
Ref.\cite{ZL_JCP}, we get an increase by a factor of $1.45$, which is
a considerable correction in the right direction. Besides the
experimental uncertainty in determination of $T_K$, we note also that
more than one pair of solitonic states might accompany the
transition. This would significantly increase the corresponding
contribution to the surface tension and the fragility.

We note that since $(T_g - T_K)/T_K \propto 1/\Delta c_p$, this
$T$-independent correction to the surface tension would be less
significant for fragile substances. ($(T_g - T_K)/T_K \lesssim 1$ for
known substances). This reflects the old notion that the temperature
dependence of $s_c$ is the leading contribution to the viscous
slowdown in a supercooled melt, except in strong liquids whose
slowdown is nearly Arrhenius-like. Incidentally, the strongest liquids
from the survey by Stevenson and Wolynes,\cite{StevensonW}, namely
GeO$_2$, and BeF$_2$, and ZnCl$_2$, do conform to the relation
$m_\sthermo = 34.7 \times \Delta c_p(T_g)$. This is consistent with
the present results, since we do not expect solitonic states in these
wide-gap materials.

\section{The charge on the coordination defects and the relation to
  earlier defect theories}
\label{sec.model}

We have argued previously\cite{ZL_JCP} that the intrinsic midgap
states should be typically charged: A half-filled defect is
essentially a neutral molecule embedded in a dielectric medium and
should be generally stabilized by adding a charge, because of ensuing
polarization.  To analyze distinct charge states of the coordination
defects in the present approach, we will use the same methodology as
in Sections \ref{parent} and \ref{soliton} above: We will consider
distinct charge states on the defects in the \emph{parent} structure,
with the usual understanding that the corresponding motif in the
actual, deformed structure will be significantly delocalized. It is
during the analysis of charged defects in the parent structure that we
will be able to make a connection between the present approach and the
much earlier specific proposals on the defect states in
chalcogenides.\cite{KAF, PhysRevLett.35.1293}

The large spatial extent of the malcoordination-based states is a
result of the lattice's attempt to mitigate the energy cost of what
would have been a \emph{local} defect in the parent structure. The
converse is also true: According to Eq.~(\ref{energyXI}), the energy
cost of localizing malcoordination would well exceed the energy
density typical of an equilibrium melt. Thus in the analogies below
between the present and earlier approaches, one should keep in mind
that on the one hand, the earlier proposals ascribing midgap states to
specific local defects can be thought of as an ultralocal limit of the
present theory. On the other hand, we see that this ultralocal limit
is somewhat artificial in that it greatly overestimates the energy cost
of the defects.

\begin{figure}[t]
\centering
\includegraphics[width= 0.7 \figurewidth]{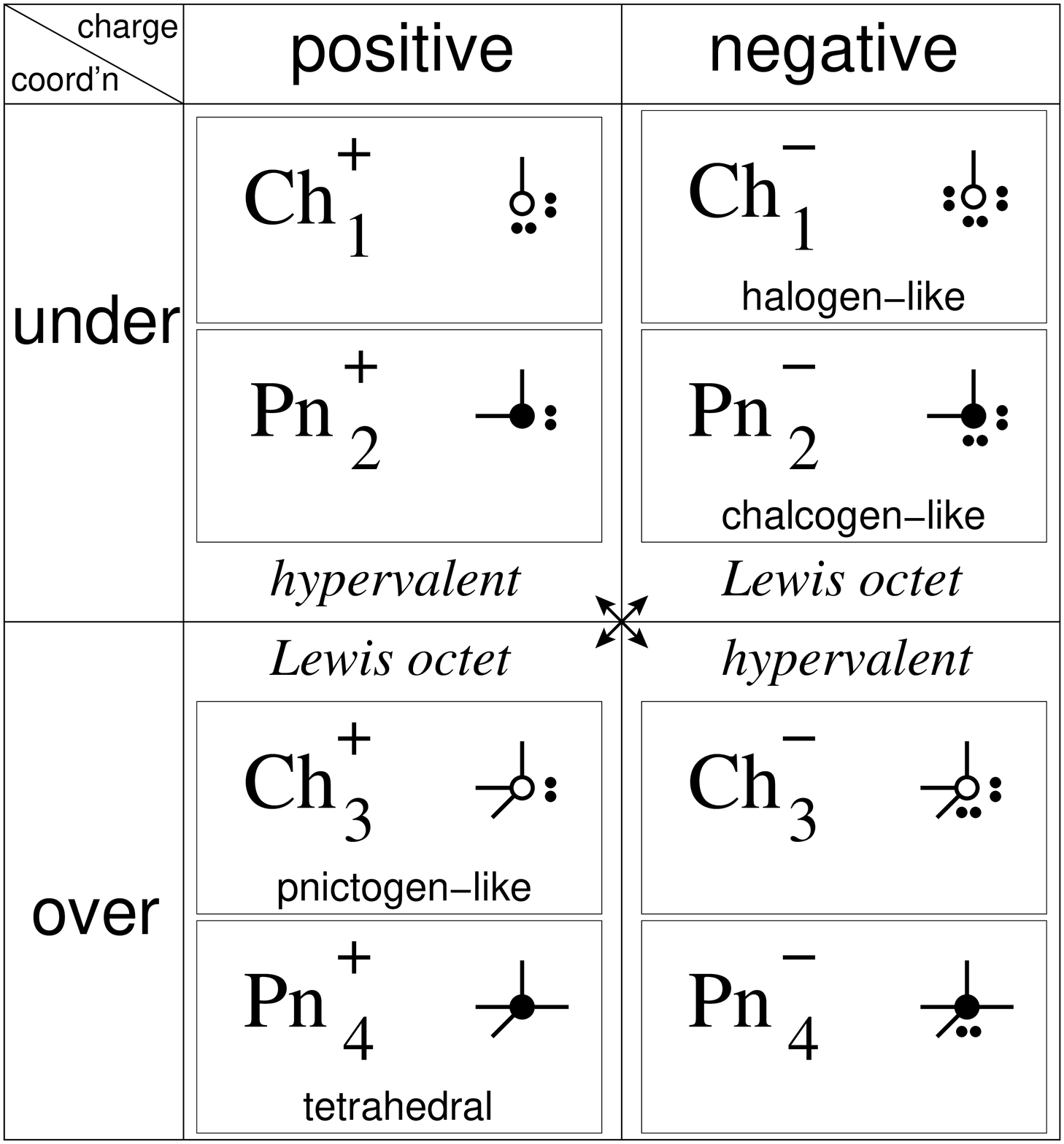}
\caption{A compilation of the possible charged states of singly
  malcoordinated atoms. Here, ``Ch''=chalcogen,
  ``Pn''=pnictogen. Neutral states, not shown, imply dangling bonds
  and would be energetically costly.}
\label{DefectsTable}
\end{figure}

Because there is only one basic type of the defect, namely a singly
under- or over-coordinated atom, and three distinct charges: -1, 0, 1,
the possible combinations of these variables are only few, see
Fig.~\ref{DefectsTable}. Already an elementary analysis yields that
charged states are expected to be stabilized, consistent with
independent arguments from Ref. \cite{ZL_JCP} Indeed, four of the
resulting configurations, namely those listed in the ``Lewis octet''
sectors, satisfy the octet rules and thus are expected to be
particularly stable. Specifically, a positively charged
overcoordinated chalcogen is chemically equivalent to a pnictogen and
a negatively charged undercoordinated pnictogen is equivalent to a
chalcogen. These two configurations are expected to stabilize the
distorted octahedral arrangement of the ambient lattice.
On the other hand, the Pn$_4^+$ configuration is unstable toward
tetrahedral order, c.f. Fig.~\ref{topstable}(c) and
Fig.~\ref{topstableAsH}, and as such, would tend to frustrate the
$pp\sigma$-network. The situation with the entries in the
``hypervalent'' and ``hypovalent'' sectors appears less certain. First
off, we have used the label ``hypervalent'' in the right bottom sector
in reference to the formal electron count around the atom exceeding 8
and the low magnitude of electronegativity variation in the compounds
in question. The T-shaped bond geometry in the Ch$_3^-$ case was
chosen by analogy with small molecules with the same electron count,
such as ClF$_3$. If the T-shaped bond geometry is indeed favored for
this type of defect, there will be an additional penalty for a
negatively charged defect to make a turn at a chalcogen.  Now, the
left upper section represents a hypovalent configuration. In the worst
case, these hypovalent configurations would favor a triangular
bipyramidal arrangement, in which there is linear alignment for at
least two bonds. We have not listed neutral defects in the
table. These would amount to having dangling bonds and thus are deemed
energetically costly. Finally note that pairs of oppositely charged
defects with complementary coordinations will either annihilate or be
particularly stable, if their annihilation is topologically
forbidden. This attraction is indicated by the double-ended arrows in
Fig.~\ref{DefectsTable}. A specific realization of an attractive,
topologically stable pair of defects in a \emph{deformed} structure is
shown in Fig.~\ref{topstableAsH}. This configuration formally
corresponds to a bound Pn$_2^-$-Pn$_4^+$ pair.

Several specific configurations from Fig.~\ref{DefectsTable},
including Ch$_3^+$, Ch$_3^-$, Ch$_1^+$, and Pn$_4^+$, have been
proposed as relevant defect species earlier, based on the putative
presence of dangling bonds\cite{KAF, PhysRevLett.35.1293} vector-type
charge-density waves,\cite{ShimoiFukutome2} and molecular dynamics
studies using energies determined by tight-binding
methods.\cite{Simdyankin} Specifically in the venerable approach of
Kastner et al.\cite{KAF}, pairs of defects, called valence-alternation
pairs (VAP), could spontaneously arise in glass because of a presumed
instability of a pair of dangling bonds toward creation of two charged
defects: 2Ch$_3^0 \rightarrow$Ch$_3^+ +$Ch$_1^-$ or 2Pn$_4^0
\rightarrow$Pn$_4^+ +$Ch$_1^-$. Despite common features with the VAP
theory, the present picture is distinct in several key aspects, in
addition to the aforementioned ones.  We have seen in Section
\ref{parent} the coordination defects are a necessary prerequisite for
molecular transport and are themselves mobile. In other words, (the
delocalized versions of) the specific configurations from
Fig.~\ref{DefectsTable} are simply transient configurations arising
during motions of the structural interfaces in a quenched melt or
frozen glass.

\section{Concluding Remarks}

We have proposed a structural model, by which the structure and
electronic anomalies of an important class of vitreous alloys emerge
in a self-consistent fashion. The model allows one to reconcile two
seemingly conflicting characteristics of quenched melts and frozen
glasses. On the one hand, these materials exhibit remarkable
thermodynamic and mechanical stability, only slightly inferior to the
corresponding crystals. On the other hand, these materials are also
multiply degenerate thus allowing for molecular transport.  The
stability of the lattice stems from its close relationship with a
highly symmetric, fully bonded structure.\cite{ZLMicro1} The lattice's
aperiodicity and the necessary presence of a thermodynamic number of
transition states for structural reconfigurations dictate that the
lattice exhibit a thermodynamic number of special structural motifs,
roughly one per several hundred atoms at $T_g$. These motifs host
midgap electronic states that are similar to solitonic states in
trans-polyacetylene.\cite{RevModPhys.60.781} The motifs correspond to
coordination defects in the \emph{parent} structure, which can be
defined entirely unambigously. In contrast, in the actual materials,
coordination is poorly defined because the interatomic distances and
bond angles are continuously distributed. The presence and the
characteristics of these defects have been established using an
exhaustive, systematic protocol, implying the classification of the
defects is complete.

The actual molecular mechanism of both the stability of semiconductor
glasses \emph{and} the mobility of the defected configurations relies
on the very special property of chalcogen- and pnictogen- containing
semiconductors, namely their fully developed $pp\sigma$-bond network.
Hereby, most atoms exhibit a distorted octahedral coordination. In
each pair of collinear bonds on such an atom, one is fully covalent
and the other is weaker, but still directional. The two bonds are
intrinsically related because they exchange electron density,
similarly to the donor-acceptor interactions. During defect transport,
the stronger and weaker bond exchange identities, resulting in bond
switching not unlike the Grotthuss mechanism of bond switching in
water.  The electronic states residing on these mobile coordination
defects are thus an intrinsic feature of transport in
$pp\sigma$-bonded melts and glasses.  In contrast, bond switching in
known tetrahedrally bonded semiconductors does not occur because the
bonds in these materials are homogeneously saturated throughout,
excluding the possibility of intrinsic states in these materials,
consistent with our earlier conclusions.\cite{ZL_JCP} Note that all
types of glassformers host the high strain interfacial regions that
separate low free energy aperiodic configurations and are intrinsic to
activated transport. Yet glassformers exhibiting distorted octahedral
coordination appear to be unique in their ability to host topological
\emph{electronic} states. The corresponding glasses are thus expected
to exhibit anomalies that we have associated with these electronic
states,\cite{ZL_JCP} including light-induced electron spin
resonance. This expectation is consistent with the variation of the
magnitude of light-induced anomalies across the Ge$_x$Se$_{1-x}$
series\cite{Salmon2007} for $1/3 < x < 1/2$, which display
coordination ranging from tetrahedral (smaller $x$) to distorted
octahedral (larger $x$). In this series, the octahedral ordering seems
to correlate with the light induced ESR and vice versa for the
tetrahedral bonding.\cite{Mollot1980}

The coordination defects exhibit topological features, such as
stability against annihilation, if there is a misalignment in the
motion of two defects. This peculiarity is related to another
topological feature of these states, which is especially transparent
in the continuum limit of Eq.~(\ref{Hel}):\cite{ZL_JCP} $\cH = - i v
\sigma_3 \prtl_x + \Delta(x) \sigma_1 + \epsilon(x) \sigma_2$. Here,
$\sigma_i$ are the Pauli matrices, while $ -i v$, $\epsilon(x)$, and
$\Delta(x)$ correspond, respectively, to the kinetic energy, local
one-particle gap and variation in electronegativity. The local gap
$\Delta(x)$ is space dependent and, in fact, switches sign at the
defect, thus corresponding to a rotation of a vector $(\Delta,
\epsilon)$. The orientation angle of this vector is the topological
phase associated with the defect. The phases of defects traveling
along different paths can not cancel, resulting in a stability against
annihilation.

Finally, what are the implications of the present results for direct
molecular modeling of amorphous chalcogenides and pnictides? Parent
structures differing only by defect locations can be used as the
initial configurations for MD simulations and may, conceivably, help
find transition state configurations corresponding to realistic
quenching rates. Yet even assuming the interactions can be estimated
accurately and efficiently, the problem is still very computationally
expensive, as highlighted by the present work: The number of
defectless parent structures, although sub-thermodynamic, is still
very very large. We have seen it is plausible that these structures
comprise the elusive Kauzmann state for these systems.

{\it Acknowledgments}: The authors thank David M. Hoffman, Thomas
A. Albright, Peter G. Wolynes, and the anonymous Reviewer for helpful
suggestions.  We gratefully acknowledge the Arnold and Mabel Beckman
Foundation Beckman Young Investigator Award, the Donors of the
American Chemical Society Petroleum Research Fund, and NSF grant
CHE-0956127 for partial support of this research.

\appendix

\section{The $pp\pi$-interaction } \label{app.pppi}

\begin{figure}[h]
\centering
\includegraphics[width=0.6 \figurewidth]{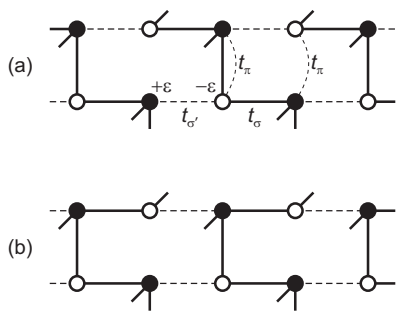}
\caption{Double chain interacting via $pp\pi$-overlaps: the
  configuration (a) is lower in energy than (b) due to
  $pp\pi$-interaction.}
\label{fig.doublechain}
\end{figure}

When the coordination is exactly octahedral, $\Pi$-shaped bonds
patterns, such as those formed by atoms 2-1-6-7 and 4-3-8-9 in
Fig.~\ref{fig.doublechain}(b), give rise to an additional energy cost
stemming from the $pp\pi$-interaction, as explained in the
following. It is partially owing to this cost, for instance, that the
layers of AsSe$_3$ pyramids in the actual structure of As$_2$Se$_3$
are so greatly puckered, see Figs. 7-9 of Ref.\cite{ZLMicro1} To
elucidate the origin and estimate the strength of the
$pp\pi$-interaction, we consider for concreteness the configurations
shown in Fig.~\ref{fig.doublechain}, periodically continued along the
double chain. Only atomic $p$-orbitals lying in the plane of the
figure are considered; these orbitals host two electrons per atom. The
nonzero entries of the Hamiltonian matrix are indicated in
Fig.~\ref{fig.doublechain}(a); the notations are from
Ref.\cite{ZLMicro1} The total energy of the configuration in
Fig.~\ref{fig.doublechain}(a) can be expanded in a series in terms of
the ratios $t'_\sn/t_\fn$ and $t_\pi/t_\fn$.  In the second order in
the expansion, the total energy is the sum of three contributions
(assuming $\epsilon=0$): The covalent bond energy is $-2t_\fn$ per
bond, the secondary bond energy equals $-(t'_\sn)^2/2t_\fn$ per bond,
and the contribution due to the $pp\pi$-interaction is given by
\begin{equation} \label{Epi} -\frac{t_\pi^2}{2t_\fn}
\end{equation}
per each vertical pair of atoms. One can similarly show that the
energy of the configuration in Fig.~\ref{fig.doublechain}(b) has the
same contributions as in (a) \emph{except} the stabilizing
contribution from Eq.~(\ref{Epi}), owing to the $\Pi$ shaped motifs.
As a result, the configuration in Fig.~\ref{fig.doublechain}(b) is
less energetically favorable. According to Eq.~(\ref{Epi}), such a
$\Pi$ motif generically costs about $0.1$ eV; typical values of the
transfer integrals can be found, e.g. in Table 1 of
Ref.\cite{ZLMicro1} This fact can be interpreted equally well either
as stabilization due to the staircase motifs in configuration (a) or
destabilization due to the $\Pi$ motifs in configuration (b).

The presence of the $pp\pi$-interaction generally affects the
energetics of defect movement, as could be seen in
Fig.~\ref{fig.doublechain}. The top chains in panels (a) and (b) can
be obtained from each other by switching all the bonds, as could be
achieved, for instance, by moving a malcoordination defect along the
chain. As a consequence, the motion of such a defect is subject to a
uniform potential. Now, each structural transition in a supercooled
melt is accompanied by the creation of two defects of opposite charge
and malcoordination.\cite{ZL_JCP} It follows from the above discussion
that in the worst case scenario, the cost of separating two defects
scales linearly with the distance, i.e. as $N^{1/3}$, where $N$ is the
size of the rearranged region. The scaling of this cost is clearly
subdominant to that of the mismatch penalty term $\gamma N^{1/2}$ from
Eq.~(\ref{FFN}).  Furthermore, at large enough $N$, it will be
energetically preferable to remove the $\Pi$ patterns separating the
original pair of defects by inserting another pair of malcoordination
defects between them. By this mechanism, which is not unlike the
mechanism of quark confinement, the cost of the $\Pi$ patterns is
limited by a fixed number which is comparable, but smaller than the
typical energy of the interface, i.e., $\gamma N^{1/2}$. We suggest
the actual effect of the $\Pi$-configurations is yet smaller. Indeed,
distinct liquid configurations should have, on the average, the same
number of $\Pi$-configurations, implying the defect movements are
subject to a zero bias, on average. As a result, the magnitude of the
overall bias for defect separation scales at most with a \emph{square
  root} of the distance, leading to a correction to the energy cost of
the reconfiguration that is proportional to $(N^{1/3})^{1/2} =
N^{1/6}$. This correction is clearly inferior to the $\gamma N^{1/2}$
term.

%


\onecolumngrid

\newpage

\begin{center} \LARGE \textbf{Supplementary Material} \end{center}

\begin{figure}[h]
\centering
\includegraphics[width=\linewidth]{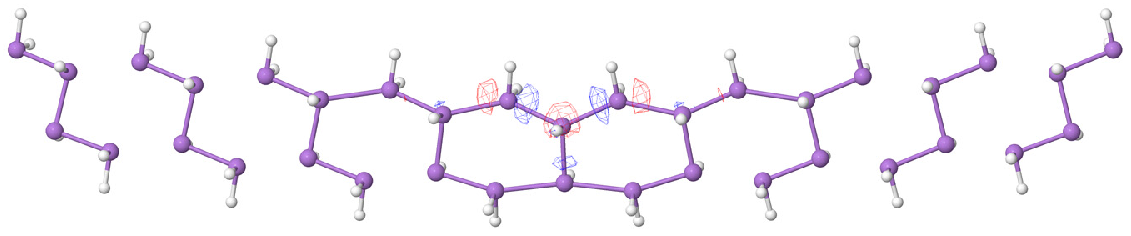}
\includegraphics[width=\linewidth]{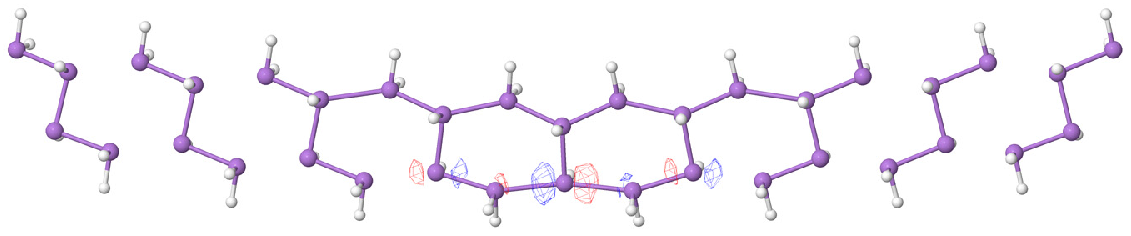}
\includegraphics[width=\linewidth]{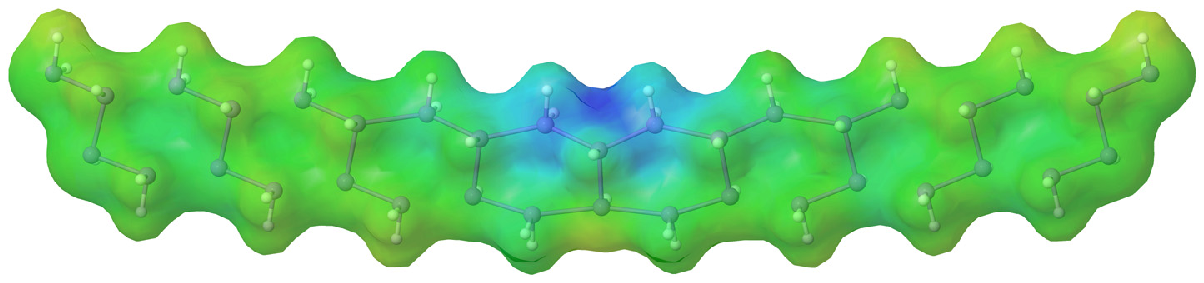}
\caption{Local charge distribution at a topologically stable pair of
  an over- and under-coordinated arsenic in a double chain of
  hydrogen-passivated arsenics from Fig. 10 of the main text. The LUMO
  and HOMO are shown in the top and middle panels respectively. The
  electrostatic potential is shown in the bottom panel, red and blue
  colors corresponding to positive and negative values respectively. }
\label{fig.dAsH2}
\end{figure}

\begin{figure}[h]
\centering
\includegraphics[width=0.7\linewidth]{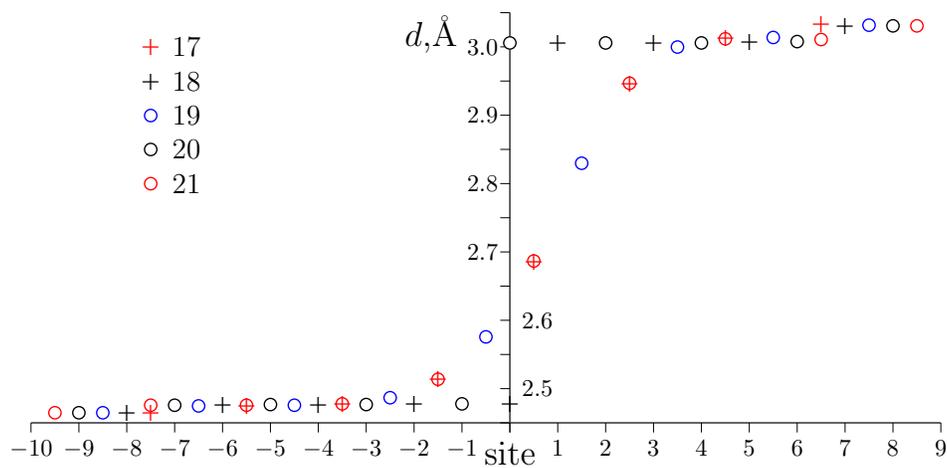}
\caption{Site dependence of As-As bond lengths in a (AsH$_2$)$_n$
   chain for several chain lengths is shown, to partially assess the
   sensitivity of the calculation to the chain length, $n$. (See the
   main text for calculational details.)  For the reader's reference we
   note that the energy gradients in the geometry-optimized chains were
   about 0.01~kcal/mol/\AA\ for even $n$, and 0.001~kcal/mol/\AA\ for odd
   $n$.}
\label{fig.AsH2_dist}
\end{figure}

\end{document}